\documentclass[nofootinbib,amsmath,prd,aps,showpacs,superscriptaddress,preprintnumbers,
tightenlines,12pt]{revtex4}
\usepackage{graphicx}
\usepackage{bm}
\usepackage{epsfig}
\usepackage{graph ics}
\usepackage{xspace}
\usepackage{color}
\usepackage{subfigure}

  \usepackage{amsmath,psfrag,subfigure,slashed,bbold}
\allowdisplaybreaks

\begin{document}

\title{The Bound-State $\beta^-$--Decay of the Neutron Revisited}

\author{A. N. Ivanov}\email{ivanov@kph.tuwien.ac.at}\affiliation{Atominstitut,
  Technische Universit\"at Wien, Stadionalle 2, A-1020 Wien, Austria}
\author{M. Pitschmann}\affiliation{Atominstitut, Technische
  Universit\"at Wien, Stadionalle 2, A-1020 Wien, Austria}
\author{N. I. Troitskaya} \affiliation{St. Petersburg State
  Polytechnical University, Polytechnicheskaya 29, 195251, Russian
  Federation} \author{Ya. A. Berdnikov} \affiliation{St. Petersburg
  State Polytechnic University, Polytechnicheskaya 29, 195251, Russian
  Federation}

\date{\today}

\begin{abstract}
This paper is addressed to the analysis of the set of observables of
the bound--state $\beta^-$--decay, which can be used for the
experimental investigation of contributions of i) interactions beyond
the Standard Model (SM) and of ii) the left--handed polarisation state
of antineutrinos. For this aim we calculate the branching ratio,
probabilities and angular distributions of probabilities of hydrogen
in the hyperfine states and of the proton--electron pair in different
spinorial states, induced by left--handed and right--handed hadronic
and leptonic currents.  The branching ratio of the bound--state
$\beta^-$--decay is calculated by taking into account radiative
corrections. We show that the probabilities of the bound--state
$\beta^-$--decay can be good observables for experimental
investigations of contributions of interactions beyond the SM, whereas
the angular distributions of probabilities are good observables for
experimental searches of contributions of the left--handed
polarisation state of antineutrinos.  
\end{abstract}

\maketitle

\section{Introduction}
\label{sec:introduction}

In Ref.~\cite{Faber:2009ts} the bound-state $\beta^-$--decay of the
neutron $n \to {\rm H} + \bar{\nu}_e$, where ${\rm H}$ is hydrogen,
has been revised by taking into account the new value of the axial
coupling constant $\lambda = - 1.2750(9)$ \cite{Abele:2008zz,
  Ivanov:2012qe} as well as new effective scalar and tensor weak
lepton--nucleon interactions. The amplitude of the continuum-state
$\beta^-$--decay of the neutron, calculated in the rest frame of the
neutron in the non--relativistic approximation for the proton,
including radiative corrections by virtual $\gamma$, $W$ and
$Z$--boson exchanges as well as QCD corrections
\cite{Sirlin:1967zza,Sirlin:1974ni,Marciano:1974tv,Sirlin:1977sv,Sirlin:1981ie,
  Marciano:1985pd,Czarnecki:2004cw,Marciano:2005ec,Czarnecki:2007th},
is given by \cite{Ivanov:2012qe}
\begin{eqnarray}\label{label1}
&& \hspace{-5mm}M(n \to p e^- \bar{\nu}_e) =
  -\,2m_n\,\frac{G_F}{\sqrt{2}}\,V_{ud}\,\Big\{[\varphi^{\dagger}_p\varphi_n][\bar{u}_e
    \gamma^0(C_V + \bar{C}_V \gamma^5) v_{\bar{\nu}_e}]\Big(1 +
  \frac{\alpha}{2\pi}\,f_{\beta^-_c}(E_e, \mu)\Big) \nonumber\\ &&-
       [\varphi^{\dagger}_p\vec{\sigma}\,\varphi_n]\cdot [\bar{u}_e
         \vec{\gamma}\,(\bar{C}_A + C_A
         \gamma^5)v_{\bar{\nu}_e}]\Big(1 +
       \frac{\alpha}{2\pi}\,f_{\beta^-_c}(E_e, \mu)\Big) +
            [\varphi^{\dagger}_p\varphi_n][\bar{u}_e (C_S + \bar{C}_S
              \gamma^5) v_{\bar{\nu}_e}]\nonumber\\ && +
            [\varphi^{\dagger}_p\vec{\sigma}\,\varphi_n]\cdot
            [\bar{u}_e \gamma^0 \vec{\gamma}\,(\bar{C}_T + C_T
              \gamma^5) v_{\bar{\nu}_e}] -
            \frac{\alpha}{2\pi}\,g_F(E_e)\,[\varphi^{\dagger}_p
              \varphi_n][\bar{u}_e\,(1 -
              \gamma^5)v_{\bar{\nu}_e}]\nonumber\\ && +
            \frac{\alpha}{2\pi}\,\lambda\, g_F(E_e)
                 [\varphi^{\dagger}_p \vec{\sigma}\,\varphi_n]\cdot
                 [\bar{u}_e \gamma^0\vec{\gamma}\,(1 -
                   \gamma^5)v_{\bar{\nu}_e}]\Big\},
\end{eqnarray}
where the coupling constants $C_j$ and $\bar{C}_j$ for $j = V$, $A$,
$S$ and $T$ describe effective weak interactions
\cite{Jackson:1957zz,Jackson1957206,Severijns:2006dr}, which may be
induced by left--handed and right--handed hadronic and leptonic
currents (see also \cite{Ivanov:2012qe}) caused by interactions beyond
the standard model (SM), for example, supersymmetric interactions
\cite{RamseyMusolf:2006vr}. They are related to the coupling
constants, analogous to those which were introduced by Herczeg
\cite{Herczeg:2001vk}, as follows (see also Eq.~(2) in Appendix G in
Ref.~\cite{Ivanov:2012qe})
\begin{eqnarray}\label{label2}
C_V &=&1 +  a^h_{LL} + a^h_{LR} + a^h_{RR} + a^h_{RL},\nonumber\\
 \bar{C}_V &=& - 1 - a^h_{LL} -
a^h_{LR} + a^h_{RR} + a^h_{RL},\nonumber\\
 C_A &=& -\lambda + a^h_{LL} - a^h_{LR} + a^h_{RR} -
a^h_{RL},\nonumber\\
\bar{C}_A &=&\lambda - a^h_{LL} + a^h_{LR} + a^h_{RR} -
a^h_{RL},\nonumber\\
C_S &=& A^h_{LL} + A^h_{LR} + A^h_{RR} +
A^h_{RL},\nonumber\\
\bar{C}_S &=& - A^h_{LL} -
A^h_{LR} + A^h_{RR} + A^h_{RL},\nonumber\\
 C_T &=& 2( \alpha^h_{LL} + \alpha^h_{RR}),\nonumber\\
 \bar{C}_T &=& 2( -
 \alpha^h_{LL} + \alpha^h_{RR}),
\end{eqnarray}
where the index $h$ means that the coupling constants are introduced
at the {\it hadronic} level \cite{Ivanov:2012qe} but not at the {\it
  quark} level as done by Herczeg \cite{Herczeg:2001vk}. In addition,
in comparison with Herczeg \cite{Herczeg:2001vk} we have taken away
the common factor $G_F V_{ud}/\sqrt2$ and defined the coupling
constants $a_{LL}^h$ and $a_{LR}^h$ as deviations from the coupling
constants of the SM.\footnote{With respect to the definition of the
  coupling constants in the original paper \cite{Herczeg:2001vk} we
  have extracted the common factor $G_FV_{ud}/\sqrt{2}$ and replaced
  $a^h_{LL}$ and $a^h_{LR}$ by $a^h_{LL} \to a^h_{LL} + (1 -
  \lambda)/2$ and $a^h_{LR} \to a^h_{LR} + (1 + \lambda)/2$,
  respectively, where the coupling constants $a^h_{LL}$ and $a^h_{LR}$
  describe deviations from the coupling constants of the SM.}  The SM
weak interactions are defined by coupling constants $C_V = - \bar{C}_V
= 1$, $C_A = - \bar{C}_A = - \lambda$ and $C_S = \bar{C}_S = C_T =
\bar{C}_T = 0$. The functions $f_{\beta^-_c}(E_e, \mu)$ and $g_F(E_e)$
in Eq.~(\ref{label1}) are equal to \cite{Ivanov:2012qe}
\begin{eqnarray}\label{label3}
f_{\beta^-_c}(E_e,\mu) &=& \frac{3}{2}\,{\ell
  n}\Big(\frac{m_p}{m_e}\Big) - \frac{11}{8} + {\ell
  n}\Big(\frac{\mu}{ m_e}\Big)\,\Big[\frac{1}{\beta}\,{\ell
    n}\Big(\frac{1 + \beta}{1 - \beta}\Big) - 2 \Big] +
\frac{1}{\beta}\,L\Big(\frac{2\beta}{1 + \beta}\Big) \nonumber\\ &&-
\frac{1}{4\beta}\,{\ell n}^2\Big(\frac{1 + \beta}{1 - \beta}\Big) +
\frac{1}{2\beta}\,{\ell n}\Big(\frac{1 + \beta}{1 - \beta}\Big) +
C_{WZ},\nonumber\\ g_F(E_e) &=& \frac{\sqrt{1 - \beta^2}}{2
  \beta}\,{\ell n}\Big(\frac{1 + \beta}{1 - \beta}\Big),
\end{eqnarray}
where $m_p$ and $m_e$ are the proton and electron masses, $\mu$ is a
finite-photon mass regularisation parameter for the regularisation of
infrared divergences in virtual one--photon exchanges
\cite{Sirlin:1967zza,Sirlin:1974ni,Marciano:1974tv,Sirlin:1977sv,Sirlin:1981ie,Marciano:1985pd,
  Czarnecki:2004cw,Marciano:2005ec,Czarnecki:2007th} (see also
\cite{Ivanov:2012qe}), $\alpha = 1/137.036$ is the fine--structure
constant \cite{Beringer:1900zz}, $\beta = \sqrt{E^2_e - m^2_e}/E_e$ is
the electron velocity and $L(2\beta/(1+\beta))$ is the Spence function
\cite{abramowitz+stegun} (see also \cite{Ivanov:2012qe}). The constant
$C_{WZ} = 10.249$ is caused by electroweak boson exchanges and QCD
corrections
\cite{Sirlin:1967zza,Sirlin:1974ni,Marciano:1974tv,Sirlin:1977sv,Sirlin:1981ie,Marciano:1985pd, Czarnecki:2004cw,Marciano:2005ec,Czarnecki:2007th}
(see also Appendix D in Ref.~\cite{Ivanov:2012qe}).

As has been pointed out in \cite{Bhattacharya:2011qm,
  Cirigliano:2009wk, Cirigliano:2012ab} (see also
\cite{Ivanov:2012qe}), the contributions of interactions beyond the SM
from the coupling constants $a^h_{LL}$ and $a^h_{LR}$ are absorbed by
a redefined axial coupling constant $\lambda_{\rm eff}$ and CKM matrix
element $(V_{ud})_{\rm eff}$ \cite{Ivanov:2012qe}, i.e.
\begin{eqnarray}\label{label4}
&&\lambda \to \lambda_{\rm eff} = \frac{\lambda -
  a^h_{LL} + a^h_{LR}}{1 + a^h_{LL} + a^h_{LR}},\nonumber\\
&&V_{ud} \to (V_{ud})_{\rm eff} = V_{ud}\,(1 + a^h_{LL} +
a^h_{LR}).
\end{eqnarray}
As has been shown in \cite{Ivanov:2012qe} the axial coupling constant
$\lambda_{\rm eff}$ is real at the level of order $10^{-4}$. After
such a redefinition the phenomenological coupling constants $C_j$ and $\bar{C}_j$
for $j = V$, $A$, $S$ and $T$ become
\begin{eqnarray}\label{label5}
C_{V,{\rm eff}} &=&1 + \frac{a^h_{RR} + a^h_{RL}}{1 + a^h_{LL} +
  a^h_{LR}}= 1 + \bar{a}^h_{RR} +
\bar{a}^h_{RL},\nonumber\\ \bar{C}_{V,{\rm eff}} &=& - 1 +
\frac{a^h_{RR} + a^h_{RL}}{1 + a^h_{LL} + a^h_{LR}} = -1 +
\bar{a}^h_{RR} + \bar{a}^h_{RL},\nonumber\\ C_{A,{\rm eff}} &=&
-\lambda_{\rm eff} + \frac{a^h_{RR} - a^h_{RL}}{1 + a^h_{LL} +
  a^h_{LR}} = -\lambda_{\rm eff} + \bar{a}^h_{RR} -
\bar{a}^h_{RL},\nonumber\\ \bar{C}_{A,{\rm eff}} &=& \lambda_{\rm eff}
+ \frac{a^h_{RR} - a^h_{RL}}{1 + a^h_{LL} + a^h_{LR}} = \lambda_{\rm
  eff} + \bar{a}^h_{RR} - \bar{a}^h_{RL},\nonumber\\ C_{S,{\rm eff}}
  &=& \frac{A^h_{LL} + A^h_{LR} + A^h_{RR} + A^h_{RL}}{1 + a^h_{LL} +
    a^h_{LR}} = \bar{A}^h_{LL} + \bar{A}^h_{LR} + \bar{A}^h_{RR} +
  \bar{A}^h_{RL},\nonumber\\ \bar{C}_{S,{\rm eff}} &=& \frac{-
    A^h_{LL} - A^h_{LR} + A^h_{RR} + A^h_{RL}}{1 + a^h_{LL} +
    a^h_{LR}} = -\bar{A}^h_{LL} - \bar{A}^h_{LR} + \bar{A}^h_{RR} +
  \bar{A}^h_{RL},\nonumber\\ C_{T,{\rm eff}} &=&
  2\,\frac{\alpha^h_{LL} + \alpha^h_{RR}}{1 + a^h_{LL} + a^h_{LR}} =
  2\,(\bar{\alpha}^h_{LL} +
  \bar{\alpha}^h_{RR}),\nonumber\\ \bar{C}_{T,{\rm eff}} &=& 2\,\frac{
    - \alpha^h_{LL} + \alpha^h_{RR}}{1 + a^h_{LL} + a^h_{LR}}=
  2\,(-\bar{\alpha}^h_{LL} + \bar{\alpha}^h_{RR}).
\end{eqnarray}
Since the velocity of the electron in the hydrogen bound state with
principal number $n$ is equal to $\beta = \alpha/n$, we take the
non-relativistic limit for the electron Dirac spinor in the
calculation of the bound-state $\beta^-$--decay of the neutron.

\section{Bound--state $\beta^-$--decay of neutron and left--handed  neutrinos}
\label{sec:lefthand}

For the calculation of the amplitude of the bound--state
$\beta^-$--decay we use the following Dirac wave function for the
antineutrino
\begin{eqnarray}\label{label6}
v_{\bar{\nu}_e} = \sqrt{E}\left(\begin{array}{c}
 \vec{\sigma}\cdot \vec{n}\,\chi_{\bar{\nu}_e}
  \\ \chi_{\bar{\nu}_e}
\end{array}\right),
\end{eqnarray}
where $\vec{n} = \vec{k}/E$ and normalisation equals
$v^{\dagger}_{\bar{\nu}_e}v_{\bar{\nu}_e} = 2 E$. For the
right--handed polarisation states of antineutrinos, corresponding to
the left--handed polarisation states of neutrinos, the Pauli wave
function $\chi_{\bar{\nu}_e}$ obeys the equation $\vec{\sigma}\cdot
\vec{n}\,\chi_{\bar{\nu}_e} = - \chi_{\bar{\nu}_e}$\footnote{The Dirac
  wave function of antineutrinos in the right--handed polarisation
  state is equal to $u_{\bar{\nu}_e} = C \bar{v}^T_{\bar{\nu}_e}$ or
  $v_{\bar{\nu}_e} = C \bar{u}^T_{\bar{\nu}_e}$, where $C = i
  \gamma^0\gamma^2$ and $T$ denotes transposition
  \cite{Itzykson:1980rh}. The wave function $u_{\bar{\nu}_e} = C
  \bar{v}^T_{\bar{\nu}_e}$ is a column matrix function with elements
  $\sqrt{E}\,(\varphi_{\bar{\nu}_e}, \vec{\sigma}\cdot
  \vec{n}\,\varphi_{\bar{\nu}_e})$, where the Pauli spinor wave
  function $\varphi_{\bar{\nu}_e}$ is equal to $\varphi_{\bar{\nu}_e}
  = - i\sigma^2 \chi^*_{\bar{\nu}_e}$ and obeys the equation
  $\vec{\sigma}\cdot \vec{n}\,\varphi_{\bar{\nu}_e} = +
  \varphi_{\bar{\nu}_e}$ \cite{Itzykson:1980rh}.}. If the axis of the
antineutrino--spin quantisation is inclined relative to the axis of
the neutron--spin quantisation with a polar angle $\vartheta$, the
Pauli wave function $\chi_{\bar{\nu}_e}$ is given by
\cite{Faber:2009ts}
\begin{eqnarray}\label{label7}
\chi_{\bar{\nu}_e} = \left(\begin{array}{c} {\displaystyle - e^{\,- i
      \varphi}\sin\frac{\vartheta}{2}} \\ {\displaystyle
    \cos\frac{\vartheta}{2}}
\end{array}\right),
\end{eqnarray}
where $\varphi$ is the azimuthal angle.  Keeping the leading order
contributions in the $\alpha$--expansion of the amplitude
Eq.~(\ref{label1}) and following \cite{Faber:2009ts}, we obtain the
amplitude of the transition $n \to {\rm H} + \bar{\nu}_e$, where
hydrogen is in the hyperfine $(ns)_F$ state with hyperfine spin $F$,
in the form
\begin{eqnarray}\label{label8}
&&M(n\to {\rm H} + \bar{\nu}_e)_{\rm rh} = G_F (V_{ud})_{\rm
    eff}\sqrt{2 m_n 2 E_{\rm H} 2 E}\,\Big(1 +
  \frac{\alpha}{2\pi}\,(f_{\beta^-_c} - 1)\Big) \,\psi^*_{(ns)_F}(0)
  \nonumber\\ &&\quad\times\Big\{(1 + g_S)\, [\varphi^{\dagger}_{p}
    \varphi_n]\,[\varphi^{\dagger}_e \chi_{_{\bar{\nu}_e}}] +
  (\lambda_{\rm eff} + g_T )\,
              [\varphi^{\dagger}_p\vec{\sigma}\,\varphi_n] \cdot
              [\varphi^{\dagger}_e\vec{\sigma}\,\chi_{_{\bar{\nu}_e}}]\Big\},
\end{eqnarray}
where the abbreviation ``rh'' means the {\it right--handed}
polarisation state. Then, $\varphi_j$ for $j = p, n, e$ and
$\chi_{\bar{\nu}_e}$ are Pauli spinorial functions of the proton,
neutron, electron and antineutrino, respectively. The term
$(-\alpha/2\pi)$ is the contribution of those terms in
Eq.~(\ref{label1}), which are proportional to
$(\alpha/2\pi)\,g_F(E_e)$. Furthermore, $(f_{\beta^-_c} - 1)$ and the
effective coupling constants $g_S$ and $g_T$ are given by
\begin{eqnarray}\label{label9}
f_{\beta^-_c} - 1 = \frac{3}{2}\,{\ell
  n}\Big(\frac{m_p}{m_e}\Big) - \frac{27}{8} + C_{WZ},
\end{eqnarray}
and
\begin{eqnarray}\label{label10}
g_S &=& \frac{1}{2}\Big((C_{S,{\rm eff}} - \bar{C}_{S,{\rm eff}}) +
(C_{V,{\rm eff}} - 1) - (\bar{C}_{V,{\rm eff}} + 1)\Big) =
\bar{A}^h_{LL} + \bar{A}^h_{LR},\nonumber\\ g_T &=& \frac{1}{2}\Big(
(C_{T,{\rm eff}} - \bar{C}_{T,{\rm eff}}) - (C_{A,{\rm eff}} +
\lambda_{\rm eff}) + (\bar{C}_{A,{\rm eff}} - \lambda_{\rm eff})\Big)
= 2 \bar{\alpha}^h_{LL}.
\end{eqnarray}
Note that the right--handed leptonic currents
with the left- and right--handed hadronic currents do not contribute
to the effective coupling constants $g_S$ and $g_T$.

Following \cite{Faber:2009ts} and keeping only the linear terms in the
expansion in powers of $g_S$ and $g_T$ we obtain the branching ratio
of the bound-state $\beta^-$--decay of the neutron
\begin{eqnarray}\label{label11}
R_{\beta^-_b} = \Big(1 + \frac{2}{1 + 3 \lambda^2_{\rm eff}}\,{\rm Re}(g_S + 3
\lambda_{\rm eff} g_T) - \Big\langle \frac{m_e}{E_e}\Big\rangle_{\rm
  SM}\,b_F\Big)\,R_{\rm SM},
\end{eqnarray}
where $\langle m_e/E_e\rangle_{\rm SM} = 0.6556$, averaged over the
electron--energy density spectrum of the neutron $\beta^-$--decay
\cite{Ivanov:2012qe}. The Fierz term $b_F$, defined to linear
approximation with respect to the Herczeg coupling constants
\cite{Ivanov:2012qe}, is given by 
\begin{eqnarray}\label{label12}
b_F = \frac{1}{1 + 3 \lambda^2_{\rm eff}}\,{\rm Re}\Big((C_{S,{\rm
    eff}} -\bar{C}_{S,{\rm eff}}) + 3\lambda_{\rm eff} (C_{T,{\rm
    eff}} - \bar{C}_{T,{\rm eff}})\Big) = \frac{2}{1 + 3
  \lambda^2_{\rm eff}}\,{\rm Re}\big(g_S + 3\lambda_{\rm eff}
g_T\big).
\end{eqnarray}
Hence, we may write the branching ratio $R$ in the form
\begin{eqnarray}\label{label13}
R_{\beta^-_b} = \Big(1 + \Big(1 -\Big\langle
\frac{m_e}{E_e}\Big\rangle_{\rm SM}\Big)\,b_F\Big)\,R_{\rm SM}.
\end{eqnarray}
As has been pointed out in \cite{Ivanov:2012qe}, the Fierz term can be
measured from the experimental data on the electron asymmetry
$A_{\exp}(E_e)$ of correlations between the neutron spin and the
electron 3--momentum and the proton--energy spectrum $a(T_p)$ (see
also \cite{Ivanov:2013fca}), related to correlations between the
3--momenta of the proton and electron. The branching ratio $R_{\rm
  SM}$, calculated in the SM, is equal to
\begin{eqnarray}\label{label14}
R_{\rm SM} = 2\pi\alpha^3\zeta(3)\,\Big(1 +
\frac{\alpha}{\pi}\,(f_{\beta^-_c} - 1)\Big)\,\frac{m_p +
  m_e}{m_n}\,\frac{E^2}{m^2_e f_n}\,\sqrt{1 +
  \frac{E^2}{(m_p + m_e)^2}} = 3.905\times 10^{-6}.
\end{eqnarray}
where $\zeta(3) = 1.202$, $E = (m^2_n - (m_p + m_e)^2)/2m_n =
0.782\,{\rm MeV}$ and $f_n= 1.755$ are the Riemann zeta function
\cite{abramowitz+stegun}, the antineutrino energy and the phase--space factor of
the neutron $\beta^-$--decay rate, including the contributions of the
corrections, caused by the ``weak magnetism'' and the proton recoil as well as
radiative corrections \cite{Ivanov:2012qe}, respectively. 

The contributions of different spinorial states to the helicity
amplitudes of the bound-state $\beta^-$--decay as functions of the
angles $\vartheta$ and $\varphi$ are given in Table I.
\begin{table}[h]
\begin{tabular}{|l|c|c|c|c|}
\hline $\sigma_n$ & $\sigma_p$ & $\sigma_e$ &
$\sigma_{\bar{\nu}_e}$& $f$\\ \hline $+\frac{1}{2} $ & $+\frac{1}{2}
$ & $-\frac{1}{2} $ &$+\frac{1}{2} $ & $(1 + g_S - \lambda_{\rm eff} -
g_T)\,\cos\frac{\vartheta}{2}$\\ \hline $+\frac{1}{2} $ &
$+\frac{1}{2} $ & $+\frac{1}{2} $ &$+\frac{1}{2} $ & $-(1 + g_S +
\lambda_{\rm eff} + g_T)\,e^{\,- i\varphi}\sin\frac{\vartheta}{2}$
\\ \hline $+\frac{1}{2} $ & $-\frac{1}{2} $ & $-\frac{1}{2} $
&$+\frac{1}{2} $ & $0$ \\ \hline $+\frac{1}{2} $ & $-\frac{1}{2} $ &
$+\frac{1}{2} $ &$+\frac{1}{2} $ & $ 2(\lambda_{\rm eff} + g_T)\,
\cos\frac{\vartheta}{2}$ \\ \hline $-\frac{1}{2} $ & $+\frac{1}{2} $ &
$-\frac{1}{2} $ &$+\frac{1}{2} $ & $ - 2 (\lambda_{\rm eff} +
g_T)\,e^{\,- i\varphi}\sin\frac{\vartheta}{2}$ \\ \hline $-\frac{1}{2}
$ & $+\frac{1}{2} $ & $+\frac{1}{2} $ &$+\frac{1}{2} $ & $0$ \\ \hline
$-\frac{1}{2} $ & $-\frac{1}{2} $ & $-\frac{1}{2} $ &$+\frac{1}{2} $ &
$(1 + g_S + \lambda_{\rm eff} + g_T)\,\cos\frac{\vartheta}{2}$
\\ \hline $-\frac{1}{2} $ & $-\frac{1}{2} $ & $+\frac{1}{2} $
&$+\frac{1}{2} $ & $-(1 + g_S - \lambda_{\rm eff} - g_T)\,e^{\,-
  i\varphi}\sin\frac{\vartheta}{2}$ \\ \hline
\end{tabular}
\caption{The contributions of different spinorial states of the
  interacting particles to the amplitudes of the bound-state
  $\beta^-$--decay of the neutron and the antineutrino in the state
  with the wave function Eq.~(\ref{label7}); $f$ is defined by $f = (1
  +
  g_S)[\varphi^{\dagger}_e\chi_{_{\bar{\nu}_e}}][\varphi^{\dagger}_p
    \varphi_n] + (\lambda_{\rm eff} + g_T)
  [\varphi^{\dagger}_e\vec{\sigma}\,\chi_{_{\bar{\nu}_e}}]\cdot [
    \varphi^{\dagger}_p\vec{\sigma}\,\varphi_n]$. }
\end{table}

Using the results in Table I we get the helicity amplitudes $M(n \to
{\rm H}_{FM_F}+ \bar{\nu}_e)_{\sigma_n,+\frac{1}{2}}$
\begin{eqnarray}\label{label15}
M(n \to {\rm H}_{00} + \bar{\nu}_e)_{+\frac{1}{2},+\frac{1}{2}} &=& +
M_0 \,\frac{1 - 3 \lambda_{\rm eff} + g_S - 3 g_T}{\sqrt{2}}\,
\cos\frac{\vartheta}{2},\nonumber\\ M(n \to {\rm H}_{1,+1} +
\bar{\nu}_e)_{+\frac{1}{2},+\frac{1}{2}} &=& - M_0 \,(1 + \lambda_{\rm
  eff} + g_S + g_T) e^{\,-
  i\varphi}\,\sin\frac{\vartheta}{2},\nonumber\\ M(n \to {\rm H}_{10}
+ \bar{\nu}_e)_{+\frac{1}{2},+\frac{1}{2}} &=& + M_0\, \frac{1 +
  \lambda_{\rm eff} + g_S + g_T}{\sqrt{2}}\,
\cos\frac{\vartheta}{2},\nonumber\\ M(n \to H_{1,-1} +
\bar{\nu}_e)_{+\frac{1}{2},+\frac{1}{2}} &=& ~0,\nonumber\\ M(n \to
H_{00} + \bar{\nu}_e)_{-\frac{1}{2},+\frac{1}{2}} &=& + M_0 \,\frac{1
  - 3 \lambda_{\rm eff} + g_S - 3 g_T }{\sqrt{2}}\, e^{\,-
  i\varphi}\sin\frac{\vartheta}{2},\nonumber\\ M(n \to {\rm H}_{1,+1}
+ \bar{\nu}_e)_{-\frac{1}{2},+\frac{1}{2}} &=&~ 0,\nonumber\\ M(n \to
{\rm H}_{10} + \bar{\nu}_e)_{-\frac{1}{2},+\frac{1}{2}} &=& - M_0
\,\frac{1 + \lambda_{\rm eff} + g_S + g_T}{\sqrt{2}}\,e^{\,-
  i\varphi}\sin\frac{\vartheta}{2},\nonumber\\ M(n \to {\rm H}_{1,-1}
+ \bar{\nu}_e)_{-\frac{1}{2},+\frac{1}{2}} &=& + M_0 \, (1 +
\lambda_{\rm eff} + g_S + g_T)\,\cos\frac{\vartheta}{2},
\end{eqnarray}
where $M_0$ is given by
\begin{eqnarray}\label{label16}
M_0 = G_F (V_{ud})_{\rm eff}\sqrt{2 m_n 2 E_{\rm H} 2 E}\,\Big(1 +
\frac{\alpha}{2\pi}\,(f_{\beta^-_c} - 1)\Big) \,\psi^*_{(ns)_F}(0).
\end{eqnarray}
Following \cite{Faber:2009ts} we define the angular distributions of the
production of hydrogen in the hyperfine states with $F = 0$ and $F =
1$ in the bound-state $\beta^-$--decay of the polarised neutron
\begin{eqnarray}\label{label17}
4\pi \frac{d W^{(\pm)}_{F = 0}(\theta)}{d \Omega} &=&
\Big\{\frac{1}{8} \frac{(1 - 3 \lambda_{\rm eff})^2}{1 + 3
  \lambda^2_{\rm eff}} + \frac{3}{4}\frac{(1 + \lambda_{\rm eff})(1 -
  3\lambda_{\rm eff})}{(1 + 3 \lambda^2_{\rm eff})^2}\,{\rm
  Re}(\lambda_{\rm eff}\,g_S - g_T) + \frac{1}{8} \frac{1}{1 + 3
  \lambda^2_{\rm eff}}\,|g_S - 3 g_T|^2\nonumber\\ &-& \frac{1}{8}
\frac{(1 - 3 \lambda_{\rm eff})}{(1 + 3 \lambda^2_{\rm eff})^2}\Big[(1
  - 3\lambda_{\rm eff})(|g_S|^2 + 3 |g_T|^2) + 4 {\rm Re}(g_S + 3
  \lambda_{\rm eff} g_T)\,{\rm Re}(g_S - 3 g_T)\Big]\nonumber\\ &+&
\frac{1}{2} \frac{(1 - 3 \lambda_{\rm eff})^2}{(1 + 3 \lambda^2_{\rm
    eff})^3}\,[{\rm Re}(g_S + 3 \lambda_{\rm eff} g_T)]^2\Big\}\,(1
\mp \cos\theta),\nonumber\\ 4\pi \frac{d W^{(\pm)}_{F = 1}(\theta)}{d
  \Omega} &=& \Big\{\frac{3}{8} \frac{(1 + \lambda_{\rm eff})^2}{1 + 3
  \lambda^2_{\rm eff}}- \frac{3}{4} \frac{(1 + \lambda_{\rm eff})(1 -
  3\lambda_{\rm eff})}{(1 + 3\lambda^2_{\rm eff})^2}\,{\rm
  Re}(\lambda_{\rm eff}\,g_S - g_T) + \frac{3}{8} \frac{1}{1 + 3
  \lambda^2_{\rm eff}}\,|g_S + g_T|^2\nonumber\\ &-& \frac{3}{8}
\frac{(1 + \lambda_{\rm eff})}{(1 + 3 \lambda^2_{\rm eff})^2}\,\Big[(1
  + \lambda_{\rm eff})\,(|g_S|^2 + 3 |g_T|^2) + 4\,{\rm Re}(g_S + 3
  \lambda_{\rm eff} g_T)\,{\rm Re}(g_S + g_T)\Big]\nonumber\\ &+&
\frac{3}{2} \frac{(1 + \lambda_{\rm eff})^2}{(1 + 3 \lambda^2_{\rm
    eff})^3}\,[{\rm Re}(g_S + 3 \lambda_{\rm eff}
  g_T)]^2\Big\}\,\Big(1 \pm \frac{1}{3}\,\cos\theta\Big).
\end{eqnarray}
The upper indices $(\pm)$ in $W^{(\pm)}_F(\theta)$ correspond to the
neutron polarisation, $\theta = \pi - \vartheta$ is the angle between
the neutron polarisation and the 3--momentum of hydrogen.

From Table I we define the angular distributions of the probabilities
$W^{(\pm)}_1(\theta)$, $W^{(\pm)}_2(\theta)$, $W^{(\pm)}_3(\theta)$
and $W^{(\pm)}_4(\theta)$ of the neutron decay into the bound $(p
e^-)$ spinorial states $|+1/2\rangle_p |+ 1/2\rangle_e$, $|+
1/2\rangle_p |- 1/2\rangle_e$, $|- 1/2\rangle_p |+ 1/2\rangle_e$ and
$|- 1/2\rangle_p |- 1/2\rangle_e$ (see \cite{Byrne:2001sj}),
respectively. For the neutron spin polarisation $\sigma_n = + 1/2$ the
angular distributions of the probabilities $W^{(+)}_1(\theta)$,
$W^{(+)}_2(\theta)$, $W^{(+)}_3(\theta)$ and $W^{(+)}_4(\theta)$ are
equal to
\begin{eqnarray}\label{label18}
4\pi \frac{d W_1^{(+)}(\theta)}{d \Omega} &=& \Big\{\frac{1}{4}
\frac{(1 + \lambda_{\rm eff})^2}{1 + 3 \lambda^2_{\rm eff}} -
\frac{1}{2} \frac{(1 + \lambda_{\rm eff})(1 - 3\lambda_{\rm eff})}{(1
  + 3 \lambda^2_{\rm eff})^2}\,{\rm Re}(\lambda_{\rm eff}\,g_S - g_T)
+ \frac{1}{4} \frac{1}{1 + 3\lambda^2_{\rm eff}}\,|g_S + g_T|^2
\nonumber\\ &-& \frac{1}{4}\,\frac{(1 + \lambda_{\rm eff})}{(1 + 3
  \lambda_{\rm eff})^2}\Big[(1 + \lambda_{\rm eff})\, (|g_S|^2 + 3
  |g_T|^2) + 4 {\rm Re}(g_S + 3\lambda_{\rm eff} g_T)\,{\rm Re}(g_S +
  g_T)\Big]\nonumber\\ &+& \frac{(1 + \lambda_{\rm eff})^2}{(1 +
  3\lambda^2_{\rm eff})^3}\,[{\rm Re}(g_S + 3\lambda_{\rm eff}
  g_T)]^2\Big\}\,(1 + \cos\theta),\nonumber\\ 4\pi \frac{d
  W_2^{(+)}(\theta)}{d \Omega} &=& \Big\{\frac{1}{4} \frac{(1 -
  \lambda_{\rm eff})^2}{1 + 3 \lambda^2_{\rm eff}} + \frac{1}{2}
\frac{(1 - \lambda_{\rm eff})(1 + 3\lambda_{\rm eff})}{(1 +
  3\lambda^2_{\rm eff})^2}\,{\rm Re}(\lambda_{\rm eff}\,g_S - g_T) +
\frac{1}{4} \frac{1}{1 + 3\lambda^2_{\rm eff}}\,|g_S - g_T|^2
\nonumber\\ &-& \frac{1}{4}\,\frac{(1 - \lambda_{\rm eff})}{(1 +
  3\lambda_{\rm eff})^2}\Big[(1 - \lambda_{\rm eff})\, (|g_S|^2 + 3
  |g_T|^2) + 4\,{\rm Re}(g_S + 3\lambda_{\rm eff} g_T)\,{\rm Re}(g_S -
  g_T)\Big]\nonumber\\ &+& \frac{(1 - \lambda_{\rm eff})^2}{(1 +
    3\lambda^2_{\rm eff})^3}\,[{\rm Re}(g_S + 3\lambda_{\rm eff}
    g_T)]^2 \Big\}\,(1 - \cos\theta),\nonumber\\ 4\pi \frac{d
    W_3^{(+)}(\theta)}{d \Omega} &=& \Big\{ \frac{\lambda^2_{\rm
      eff}}{1 + 3 \lambda^2_{\rm eff}}- \frac{2 \lambda_{\rm eff}}{(1
    + 3\lambda^2_{\rm eff})^2}\,{\rm Re}(\lambda_{\rm eff}\,g_S -
  g_T) + \frac{1}{1 + 3 \lambda^2_{\rm
        eff}}\,|g_T|^2\nonumber\\ &-&
  \frac{\lambda_{\rm eff}}{(1 + 3\lambda^2_{\rm
        eff})^2}\Big[\lambda_{\rm eff}(|g_S|^2 + 3 |g_T|^2) + 4 {\rm
        Re}(g_S + 3\lambda_{\rm eff} g_T)\,{\rm
        Re}(g_T)\Big]\nonumber\\ &+& \frac{4\lambda^2_{\rm
        eff}}{(1 + 3\lambda^2_{\rm eff})^3}\,[{\rm Re}(g_S +
      3\lambda_{\rm eff} g_T)]^2\Big\}\, (1 -
  \cos\theta),\nonumber\\ 4\pi \frac{d W_4^{(+)}(\theta)}{d \Omega}
  &=& 0,
\end{eqnarray}
For the neutron spin polarisation $\sigma_n = -1/2$ the angular
distributions of the probabilities $W^{(-)}_1(\theta)$,
$W^{(-)}_2(\theta)$, $W^{(-)}_3(\theta)$ and $W^{(-)}_4(\theta)$
are
\begin{eqnarray}\label{label19}
\frac{d W^{(-)}_1(\theta)}{d \Omega} &=& \frac{d
  W^{(+)}_4(\pi - \theta)}{d \Omega},\nonumber\\ \frac{d
  W^{(-)}_2(\theta)}{d \Omega} &=& \frac{d W^{(+)}_3(\pi -
  \theta)}{d \Omega},\nonumber\\ \frac{d
  W^{(-)}_3(\theta)}{d \Omega} &=& \frac{d W^{(+)}_2(\pi -
  \theta)}{d \Omega},\nonumber\\ \frac{d
  W^{(-)}_4(\theta)}{d \Omega} &=& \frac{d W^{(+)}_1(\pi -
  \theta)}{d \Omega}.
\end{eqnarray}
In comparison with \cite{Faber:2009ts} we have expanded the angular
distributions up to second order in scalar and tensor couplings. 
We have found that to linear approximation with
respect to the coupling constants $g_S$ and $g_T$ the bound--state $\beta^-$--decay of the free neutron is
sensitive to the effective coupling constant ${\rm Re}(\lambda_{\rm
  eff}\,g_S - g_T)$ only. The terms of order $|g_S|^2$, $|g_T|^2$ and
${\rm Re}(g^*_Sg_T)$ contain more complicated combinations of the
coupling constants $g_S$ and $g_T$.  However, both of these contributions
contain no information on right--handed leptonic currents.

\section{Bound--state $\beta^-$--decay of neutron and right--handed neutrinos}
\label{sec:righthand}

The absence of contributions of right--handed leptonic currents in the
probabilities of the bound--state $\beta^-$--decay, calculated in
section \ref{sec:lefthand}, is not a surprise, because we have used
there the wave functions of antineutrinos in the right--handed
polarisation state, corresponding to the left--handed polarisation
state of neutrinos. Being multiplied by the projection operator $P_R =
(1 + \gamma^5)/2$, appearing in the right--handed leptonic currents,
the wave function of antineutrinos $v_{\bar{\nu}_e}$ in the
right--handed polarisation state with $\vec{\sigma}\cdot \vec{n}\,
\chi_{\bar{\nu}_e} = - \chi_{\bar{\nu}_e}$ gives a vanishing
contribution, i.e. $P_R v_{\bar{\nu}_e} = 0$.

In this section, we assume that antineutrinos can have also
left--handed polarisation state that is possible if antineutrinos
(neutrinos) are massive. According to \cite{Beringer:1900zz}, a mass
of the electron neutrino (antineutrino) should not exceed a few ${\rm
  eV}$. Since in the bound--state $\beta^-$--decay $E = Q_{\beta^-_c}
= 0.782\,{\rm MeV}$ \cite{Faber:2009ts}, one can neglect the
antineutrino mass with respect to the antineutrino energy $E$ and use
the Dirac wave function Eq.~(\ref{label6}). However, the Pauli wave
function $\chi_{\bar{\nu}_e}$ of antineutrinos in the left--handed
polarisation state should obey the equation $\vec{\sigma}\cdot
\vec{n}\,\chi_{\bar{\nu}_e} = + \chi_{\bar{\nu}_e}$\footnote{The
  Dirac wave function of antineutrinos in the left--handed
  polarisation state is equal to $u_{\bar{\nu}_e} = C
  \bar{v}^T_{\bar{\nu}_e}$, where the Pauli spinor wave function
  $\varphi_{\bar{\nu}_e} = - i\sigma^2 \chi^*_{\bar{\nu}_e}$ obeys the
  equation $\vec{\sigma}\cdot \vec{n}\,\varphi_{\bar{\nu}_e} = -
  \varphi_{\bar{\nu}_e}$ \cite{Itzykson:1980rh}.}. If the axis of the
antineutrino--spin quantisation is inclined relative to the axis of
the neutron--spin quantisation with a polar angle $\vartheta$, the
Pauli wave function $\chi_{\bar{\nu}_e}$ is given by
\begin{eqnarray}\label{label20}
\chi_{\bar{\nu}_e} = \left(\begin{array}{c} {\displaystyle
    \cos\frac{\vartheta}{2}} \\ {\displaystyle e^{\,+ i
      \varphi}\sin\frac{\vartheta}{2}}
\end{array}\right).
\end{eqnarray}
The amplitude of the bound--state $\beta^-$--decay of the neutron with
antineutrinos in the left--handed polarisation state is given by
\begin{eqnarray}\label{label21}
&&M(n\to {\rm H} + \bar{\nu}_e)_{\rm lh} = G_F (V_{ud})_{\rm
    eff}\sqrt{2 m_n 2 E_{\rm H} 2 E}\,\Big(1 +
  \frac{\alpha}{2\pi}\,f_{\beta^-_c} \Big)
  \nonumber\\ &&\quad\times\Big\{\bar{g}_S\, [\varphi^{\dagger}_{p}
    \varphi_n]\,[\varphi^{\dagger}_e \chi_{_{\bar{\nu}_e}}] +
  \bar{g}_T\, [\varphi^{\dagger}_p\vec{\sigma}\,\varphi_n] \cdot
      [\varphi^{\dagger}_e\vec{\sigma}\,\chi_{_{\bar{\nu}_e}}]
      \Big\}\,\psi^*_{(ns)_F}(0),
\end{eqnarray}
where the abbreviation ``lh'' means the {\it left--handed}
  polarisation state. The coupling constants $\bar{g}_S$ and
$\bar{g}_T$ are equal to
\begin{eqnarray}\label{label22}
\bar{g}_S &=& \frac{1}{2}\Big((C_{S,{\rm eff}} + \bar{C}_{S,{\rm
    eff}}) + (C_{V,{\rm eff}} + \bar{C}_{V,{\rm eff}})\Big) =
\bar{A}^h_{RR} + \bar{A}^h_{RL} + \bar{a}^h_{RR} +
\bar{a}^h_{RL},\nonumber\\ \bar{g}_T &=& \frac{1}{2}\Big( (C_{T,{\rm
    eff}} + \bar{C}_{T,{\rm eff}}) - (C_{A,{\rm eff}} +
\bar{C}_{A,{\rm eff}})\Big) = 2 \bar{\alpha}^h_{RR} - \bar{a}^h_{RR} +
\bar{a}^h_{RL}.
\end{eqnarray}
One may see that the coupling constants $\bar{g}_S$ and $\bar{g}_T$
are defined in terms of the contributions of the right--handed
leptonic currents and left(right)--handed hadronic currents only. The
coupling constants $\bar{a}^h_{RR}$ and $\bar{a}^h_{RL}$ can in
principle be induced by exchanges of electroweak $W^{\pm}_R$--bosons,
causing effective low--energy current--current interactions $(V +
A)_{\rm leptonic}(V + A)_{\rm hadronic}$ and $(V + A)_{\rm leptonic}
(V - A)_{\rm hadronic}$
\cite{Beg:1977ti,Holstein:1977qn,Carnoy:1988uv}. Of course, the
contributions of these interactions can be screened by scalar and
tensor interactions with coupling constants $\bar{A}^h_{RR}$,
$\bar{A}^h_{RL}$ and $2 \bar{\alpha}^h_{RR}$.

The contributions of different spinorial states to the helicity
amplitudes of the bound--state $\beta^-$--decay of the neutron with
the antineutrino in the left--handed polarisation state as functions
of the angles $\vartheta$ and $\varphi$ are given in Table II.
\begin{table}[h]
\begin{tabular}{|l|c|c|c|c|}
\hline $\sigma_n$ & $\sigma_p$ & $\sigma_e$ & $\sigma_{\bar{\nu}_e}$&
$f$\\ \hline $+\frac{1}{2} $ & $+\frac{1}{2} $ & $-\frac{1}{2} $
&$-\frac{1}{2} $ & $(\bar{g}_S - \bar{g}_T) \,e^{\,+i
  \varphi}\,\sin\frac{\vartheta}{2}$\\ \hline $+\frac{1}{2} $ &
$+\frac{1}{2} $ & $+\frac{1}{2} $ &$-\frac{1}{2} $ & $(\bar{g}_S +
\bar{g}_T) \cos\frac{\vartheta}{2}$\\ \hline $+\frac{1}{2} $ &
$-\frac{1}{2} $ & $-\frac{1}{2} $ &$-\frac{1}{2} $ & $0$ \\ \hline
$+\frac{1}{2} $ & $-\frac{1}{2} $ & $+\frac{1}{2} $ &$-\frac{1}{2} $ &
$ 2 \bar{g}_T \,e^{\,+i \varphi}\,\sin\frac{\vartheta}{2}$ \\ \hline
$-\frac{1}{2} $ & $+\frac{1}{2} $ & $-\frac{1}{2} $ &$-\frac{1}{2} $ &
$2 \bar{g}_T \cos\frac{\vartheta}{2}$ \\ \hline $-\frac{1}{2} $ &
$+\frac{1}{2} $ & $+\frac{1}{2} $ &$-\frac{1}{2} $ & $0$ \\ \hline
$-\frac{1}{2} $ & $-\frac{1}{2} $ & $-\frac{1}{2} $ &$-\frac{1}{2} $ &
$(\bar{g}_S + \bar{g}_T) \,e^{\,+i \varphi}\,\sin\frac{\vartheta}{2}$
\\ \hline $-\frac{1}{2} $ & $-\frac{1}{2} $ & $+\frac{1}{2} $
&$-\frac{1}{2} $ & $(\bar{g}_S - \bar{g}_T) \cos\frac{\vartheta}{2}$
\\ \hline
\end{tabular}
\caption{The contributions of different spinorial states of the
  interacting particles to the amplitudes of the bound-state
  $\beta^-$--decay of the neutron and the antineutrino in the state
  with the wave function Eq.~(\ref{label20}); $f$ is defined by $f =
  \bar{g}_S\,[\varphi^{\dagger}_e\chi_{_{\bar{\nu}_e}}][\varphi^{\dagger}_p
    \varphi_n] + \bar{g}_T\,
      [\varphi^{\dagger}_e\vec{\sigma}\,\chi_{_{\bar{\nu}_e}}]\cdot
      [ \varphi^{\dagger}_p\vec{\sigma}\,\varphi_n]$. }
\end{table}

Using the results in Table II we get the helicity amplitudes $M(n \to
{\rm H}_{F M_F}+ \bar{\nu}_e)_{\sigma_n,-\frac{1}{2}}$
\begin{eqnarray}\label{label23}
M(n \to {\rm H}_{00} + \bar{\nu}_e)_{+\frac{1}{2},-\frac{1}{2}} &=& +
M_0 \,\frac{\bar{g}_S - 3 \bar{g}_T}{\sqrt{2}}\, e^{\, +
  i\varphi}\,\sin\frac{\vartheta}{2},\nonumber\\ M(n \to {\rm
  H}_{1,+1} + \bar{\nu}_e)_{+\frac{1}{2},-\frac{1}{2}} &=& + M_0
\,(\bar{g}_S + \bar{g}_T) \,\cos\frac{\vartheta}{2},\nonumber\\ M(n
\to {\rm H}_{10} + \tilde{\nu}_e)_{+\frac{1}{2},-\frac{1}{2}} &=& +
M_0\, \frac{\bar{g}_S + \bar{g}_T}{\sqrt{2}}\, e^{\, +
  i\varphi}\,\sin\frac{\vartheta}{2},\nonumber\\ M(n \to {\rm
  H}_{1,-1} + \bar{\nu}_e)_{+\frac{1}{2},-\frac{1}{2}} &=&
~0,\nonumber\\ M(n \to {\rm H}_{00} +
\bar{\nu}_e)_{-\frac{1}{2},-\frac{1}{2}} &=& - M_0 \,\frac{\bar{g}_S -
  3 \bar{g}_T}{\sqrt{2}}\, \cos\frac{\vartheta}{2},\nonumber\\ M(n \to
    {\rm H}_{1,+1} + \bar{\nu}_e)_{-\frac{1}{2},-\frac{1}{2}} &=&~
    0,\nonumber\\ M(n \to {\rm H}_{10} +
    \bar{\nu}_e)_{-\frac{1}{2},-\frac{1}{2}} &=& + M_0\,
    \frac{\bar{g}_S +
      \bar{g}_T}{\sqrt{2}}\,\cos\frac{\vartheta}{2},\nonumber\\ M(n
    \to {\rm H}_{1,-1} + \bar{\nu}_e)_{-\frac{1}{2},-\frac{1}{2}} &=&
    + M_0\,(\bar{g}_S + \bar{g}_T)\,e^{\, +
      i\varphi}\,\sin\frac{\vartheta}{2}.
\end{eqnarray}
Taking into account the contributions of the {\it left--handed}
polarisation states of antineutrinos for the angular distributions we
obtain the following expressions
\begin{eqnarray}\label{label24}
&&4\pi \frac{d W^{(\pm)}_{F = 0}(\theta)}{d \Omega} =\Big\{\frac{1}{8}
  \frac{(1 - 3 \lambda_{\rm eff})^2}{1 + 3 \lambda^2_{\rm eff}} +
  \frac{3}{4} \frac{(1 + \lambda_{\rm eff})(1 - 3\lambda_{\rm
      eff})}{(1 + 3 \lambda^2_{\rm eff})^2}\,{\rm Re}(\lambda_{\rm
    eff}\,g_S - g_T) + \frac{1}{8} \frac{1}{1 + 3 \lambda^2_{\rm
      eff}}\,|g_S - 3 g_T|^2\nonumber\\ &-& \frac{1}{8} \frac{(1 - 3
    \lambda_{\rm eff})}{(1 + 3 \lambda^2_{\rm eff})^2}\Big[(1 -
    3\lambda_{\rm eff})(|g_S|^2 + 3 |g_T|^2 + |\bar{g}_S|^2 + 3
    |\bar{g}_T|^2) + 4 {\rm Re}(g_S + 3 \lambda_{\rm eff} g_T)\,{\rm
      Re}(g_S - 3 g_T)\Big]\nonumber\\ &+& \frac{1}{2} \frac{(1 - 3
    \lambda_{\rm eff})^2}{(1 + 3 \lambda^2_{\rm eff})^3}\,[{\rm
      Re}(g_S + 3 \lambda_{\rm eff} g_T)]^2\Big\}\,(1 \mp \cos\theta)
  + \frac{1}{8} \frac{1}{1 + 3\lambda^2_{\rm eff}}\,|\bar{g}_S - 3
  \bar{g}_T|^2\,(1 \pm \cos\theta),\nonumber\\ &&4\pi \frac{d
    W^{(\pm)}_{F = 1}(\theta)}{d \Omega} = \Big\{\frac{3}{8} \frac{(1
    + \lambda_{\rm eff})^2}{1 + 3 \lambda^2_{\rm eff}}- \frac{3}{4}
  \frac{(1 + \lambda_{\rm eff})(1 - 3\lambda_{\rm eff})}{(1 +
    3\lambda^2_{\rm eff})^2}\,{\rm Re}(\lambda_{\rm eff}\,g_S - g_T) +
  \frac{3}{8} \frac{1}{1 + 3 \lambda^2_{\rm eff}}\,|g_S +
  g_T|^2\nonumber\\ &-& \frac{3}{8} \frac{(1 + \lambda_{\rm eff})}{(1
    + 3 \lambda^2_{\rm eff})^2}\,\Big[(1 + \lambda_{\rm
      eff})\,(|g_S|^2 + 3 |g_T|^2 + |\bar{g}_S|^2 + 3 |\bar{g}_T|^2) +
    4\,{\rm Re}(g_S + 3 \lambda_{\rm eff} g_T)\,{\rm Re}(g_S +
    g_T)\Big]\nonumber\\ &+& \frac{3}{2} \frac{(1 + \lambda_{\rm
      eff})^2}{(1 + 3 \lambda^2_{\rm eff})^3}\,[{\rm Re}(g_S + 3
    \lambda_{\rm eff} g_T)]^2\Big\}\,\Big(1 \pm
  \frac{1}{3}\,\cos\theta\Big) + \frac{3}{8} \frac{1}{1 + 3
    \lambda^2_{\rm eff}}\,|\bar{g}_S + \bar{g}_T|^2\,\Big(1 \mp
  \frac{1}{3}\,\cos\theta\Big)
\end{eqnarray}
and
\begin{eqnarray}\label{label25}
4\pi \frac{d W_1^{(+)}(\theta)}{d \Omega} &=&
\Big\{\frac{1}{4}\,\frac{(1 + \lambda_{\rm eff})^2}{1 + 3
  \lambda^2_{\rm eff}} - \frac{1}{2}\,\frac{(1 + \lambda_{\rm eff})(1
  - 3\lambda_{\rm eff})}{(1 + 3 \lambda^2_{\rm eff})^2}\,{\rm
  Re}(\lambda_{\rm eff}\,g_S - g_T) + \frac{1}{4}\,\frac{1}{1 +
  3\lambda^2_{\rm eff}}\,|g_S + g_T|^2 \nonumber\\ &-&
\frac{1}{4}\,\frac{(1 + \lambda_{\rm eff})}{(1 + 3 \lambda_{\rm
    eff})^2}\Big[(1 + \lambda_{\rm eff})\, (|g_S|^2 + 3 |g_T|^2 +
  |\bar{g}_S|^2 + 3 |\bar{g}_T|^2) + 4\,{\rm Re}(g_S + 3\lambda_{\rm
    eff} g_T)\,{\rm Re}(g_S + g_T)\Big]\nonumber\\ &+& \frac{(1 +
  \lambda_{\rm eff})^2}{(1 + 3\lambda^2_{\rm eff})^3}\,[{\rm Re}(g_S +
  3\lambda_{\rm eff} g_T)]^2\Big\}\,(1 + \cos\theta) +
\frac{1}{4}\,\frac{1}{1 + 3\lambda^2_{\rm eff}}\,|\bar{g}_S +
\bar{g}_T|^2\,(1 - \cos\theta),\nonumber\\ 4\pi \frac{d
  W_2^{(+)}(\theta)}{d \Omega} &=& \Big\{\frac{1}{4}\, \frac{(1 -
  \lambda_{\rm eff})^2}{1 + 3 \lambda^2_{\rm eff}} +
\frac{1}{2}\,\frac{(1 - \lambda_{\rm eff})(1 + 3\lambda_{\rm eff})}{(1
  + 3\lambda^2_{\rm eff})^2}\,{\rm Re}(\lambda_{\rm eff}\,g_S - g_T) +
\frac{1}{4}\,\frac{1}{1 + 3\lambda^2_{\rm eff}}\,|g_S - g_T|^2
\nonumber\\ &-& \frac{1}{4} \frac{(1 - \lambda_{\rm eff})}{(1 +
  3\lambda_{\rm eff})^2}\Big[(1 - \lambda_{\rm eff})\, (|g_S|^2 + 3
  |g_T|^2 + |\bar{g}_S|^2 + 3 |\bar{g}_T|^2) + 4\,{\rm Re}(g_S +
  3\lambda_{\rm eff} g_T)\,{\rm Re}(g_S - g_T)\Big] \nonumber\\ &+&
\frac{(1 - \lambda_{\rm eff})^2}{(1 + 3\lambda^2_{\rm eff})^3}\,[{\rm
    Re}(g_S + 3\lambda_{\rm eff} g_T)]^2 \Big\}\,(1 - \cos\theta) +
\frac{1}{4}\,\frac{1}{1 + 3\lambda^2_{\rm eff}}\, |\bar{g}_S -
\bar{g}_T|^2\,(1 + \cos\theta),\nonumber\\ 4\pi \frac{d
  W_3^{(+)}(\theta)}{d \Omega} &=& \Big\{ \frac{\lambda^2_{\rm eff}}{1
  + 3 \lambda^2_{\rm eff}}- \frac{2 \lambda_{\rm eff}}{(1 +
  3\lambda^2_{\rm eff})^2}\,{\rm Re}(\lambda_{\rm eff}\,g_S - g_T) +
\frac{1}{1 + 3 \lambda^2_{\rm eff}}\,|g_T|^2 \nonumber\\ &-&
\frac{\lambda_{\rm eff}}{(1 + 3\lambda^2_{\rm
    eff})^2}\Big[\lambda_{\rm eff}(|g_S|^2 + 3 |g_T|^2 + |\bar{g}_S|^2
  + 3 |\bar{g}_T|^2) + 4\,{\rm Re}(g_S + 3\lambda_{\rm eff} g_T)\,{\rm
    Re}(g_T)\Big] \nonumber\\ &+& \frac{4\lambda^2_{\rm eff}}{(1 +
  3\lambda^2_{\rm eff})^3}\,[{\rm Re}(g_S + 3\lambda_{\rm eff} g_T)]^2
\Big\}\, (1 - \cos\theta) + \frac{1}{1 + 3 \lambda^2_{\rm
    eff}}\,|\bar{g}_T|^2\,(1 + \cos\theta) ,\nonumber\\ 4\pi \frac{d
  W_4^{(+)}(\theta)}{d \Omega} &=& 0.
\end{eqnarray}
The angular distributions of the probabilities $W^{(-)}_1(\theta)$,
$W^{(-)}_2(\theta)$, $W^{(-)}_3(\theta)$ and $W^{(-)}_4(\theta)$ of
the neutron bound $\beta^-$--decay with the neutron in the spin
polarisation state $\sigma_n = - \textstyle \frac{1}{2}$ are given by
\begin{eqnarray}\label{label26}
\frac{d W^{(-)}_1(\theta)}{d \Omega} &=& \frac{d W^{(+)}_4(\pi -
  \theta)}{d \Omega},\nonumber\\\frac{d W^{(-)}_2(\theta)}{d \Omega}
&=& \frac{d W^{(+)}_3(\pi - \theta)}{d \Omega}, \nonumber\\ \frac{d
  W^{(-)}_3(\theta)}{d \Omega} &=& \frac{d W^{(+)}_2(\pi - \theta)}{d
  \Omega},\nonumber\\\frac{d W^{(-)}_4(\theta)}{d \Omega} &=& \frac{d
  W^{(+)}_1(\pi - \theta)}{d \Omega}.
\end{eqnarray}
For the calculation of the angular distributions Eqs.(\ref{label24}),
(\ref{label25}) and (\ref{label26}) we have neglected the
contributions of order $\alpha\,|\bar{g}_S|^2/\pi$,
$\alpha\,|\bar{g}_T|^2/\pi$ and $\alpha\,{\rm Re}(\bar{g}^*_S
\bar{g}_T)/\pi$.

\section{Conclusion}
\label{sec:conclusion}

We have revised the bound-state $\beta^-$--decay of the free neutron
and addressed the paper to the analysis of the set of observables,
which can be used for experimental searches of contributions of
interactions beyond the SM and the left--handed polarisation state of
antineutrinos.  For this aim we have analysed the dependence of the
probabilities of the bound-state $\beta^-$--decay of the neutron and
their angular distributions on phenomenological coupling constants
\cite{Jackson:1957zz,Jackson1957206,Severijns:2006dr, Herczeg:2001vk},
describing the most general weak effective lepton--nucleon
interactions.

In comparison with the results, obtained in \cite{Faber:2009ts}: i) we
have analysed the probabilities of the bound--state $\beta^-$--decay
of the neutron by using the most general phenomenological
lepton--baryon weak interaction for the neutron $\beta^-$--decay, ii)
we have shown that antineutrinos in the right--handed polarisation
state interact only with the scalar and tensor left(right)--handed
baryon currents, iii) we have found that to linear approximation with
respect to Herczeg's phenomenological coupling constants, introduced
at the hadronic level \cite{Ivanov:2012qe}, these probabilities depend
only on the effective coupling constant ${\rm Re}(\lambda_{\rm eff}g_S
- g_T)$, which carries no information about weak interactions, caused
by right--handed leptonic and left(right)--handed hadronic currents,
iv) we have calculated the contributions of the left--handed
polarisation state of antineutrinos and found that the effective
scalar $\bar{g}_S$ and tensor $\bar{g}_T$ coupling constants are
caused by contributions of the vector and axial--vector baryon
currents and left(right)--handed baryon scalar and tensor currents of
interactions beyond the SM, v) we have calculated the probabilities of
the production of hydrogen in the hyperfine states with $F = 0$ and $F
= 1$ and the probabilities of the production of the $(p e^-)$ pairs in
the spinorial states $|+1/2\rangle_p |+ 1/2\rangle_e$, $|+
1/2\rangle_p |- 1/2\rangle_e$, $|- 1/2\rangle_p |+ 1/2\rangle_e$ and
$|- 1/2\rangle_p |- 1/2\rangle_e$ and their angular distributions to
second order of the effective coupling constants of interactions
beyond the SM for antineutrinos in the right(left)--handed
polarisation state, vi) we have found that in the probabilities under
consideration the contributions of the left--handed polarisation state
of antineutrinos can be screened by the contributions of their
right--handed polarisation state. This implies that the probabilities
of the bound-state $\beta^-$--decay $W^{(\pm)}_F$, $W^{(\pm)}_1$,
$W^{(\pm)}_2$, $W^{(\pm)}_3$ and $W^{(\pm)}_4$ can be used as
observables for experimental investigations of contributions of
interactions beyond the SM without specification of the contributions
of the polarisation states of antineutrinos and vii) we have shown
that the angular distributions of most of these probabilities are good
observables for the detection of contributions of the left--handed
polarisation state of antineutrinos.

Indeed, for example, for the neutron in the spin state $\sigma_n = +
\textstyle \frac{1}{2}$ the angular distributions of the probabilities
of the production of hydrogen in the hyperfine state with $F = 0$ and
of the $(p e^-)$ pairs in the spinorial states $|+1/2\rangle_p |+
1/2\rangle_e$, $|+ 1/2\rangle_p |- 1/2\rangle_e$ and $|- 1/2\rangle_p
|+ 1/2\rangle_e$ at a fixed angle $\theta = 0$ and $\theta = \pi$ we
obtain
\begin{eqnarray}\label{label27}
4\pi \frac{d W^{(+)}_{F = 0}(\theta)}{d \Omega}\Big|_{\theta = 0} &=&
\frac{1}{4}\,\frac{1}{1 + 3\lambda^2_{\rm eff}}\, |\bar{g}_S - 3
\bar{g}_T|^2 = ~4.25\times 10^{-2}\,|\bar{g}_S - 3
\bar{g}_T|^2,\nonumber\\ 4\pi \frac{d W^{(+)}_1(\theta)}{d
  \Omega}\Big|_{\theta = \pi} &=& \frac{1}{2}\,\frac{1}{1 +
  3\lambda^2_{\rm eff}}\,|\bar{g}_S +
  \bar{g}_T|^2 = ~8.50\times 10^{-2}\,|\bar{g}_S +
\bar{g}_T|^2,\nonumber\\ 4\pi \frac{d W^{(+)}_2(\theta)}{d
  \Omega}\Big|_{\theta = 0} &=& \frac{1}{2}\,\frac{1}{1 + 3\lambda^2_{\rm eff}}\,|\bar{g}_S -
  \bar{g}_T|^2 = ~8.50\times
10^{-2}\,|\bar{g}_S - \bar{g}_T|^2,\nonumber\\ 4\pi \frac{d
  W^{(+)}_3(\theta)}{d \Omega}\Big|_{\theta = 0} &=& ~
2\,\frac{1}{1 + 3\lambda^2_{\rm eff}}~|\bar{g}_T|^2 = 34.03 \times
10^{-2}\,|\bar{g}_T|^2.
\end{eqnarray}
The numerical coefficients are calculated for $\lambda_{\rm eff} = -
1.2750$ \cite{Ivanov:2012qe}.

The probability $W^{(\pm)}_{F = 1}$ and its angular distribution
$dW^{(\pm)}_{F = 1}(\theta)/d\Omega$ are not sensitive to the
contributions of the right(left)--handed polarisation states of
antineutrinos, caused by interactions beyond the SM. Then, the
probability $W^{(+)}_4$ of the production of the $(p e^-)$ pair in the
spinorial state $|- 1/2\rangle_p |- 1/2\rangle_e$ and its angular
distribution $dW^{(+)}_4(\theta)/d\Omega$ vanish for all polarisation
states of antineutrinos. For the neutron in the spin state with
$\sigma_n = - \frac{1}{2}$ the angular distributions of the
corresponding probabilities are given in Eq.(\ref{label26}).

It is important to notice that the effective scalar $\bar{g}_S$ and
tensor $\bar{g}_T$ coupling constants, which can be measured from the
angular distributions Eq.(\ref{label27}), depend on the coupling
constants of lepton interactions with vector and axial--vector baryon
currents, caused by right--handed electroweak bosons, and the
left(right)--handed scalar and tensor baryon currents (see
Eq.(\ref{label22})).  In this case it seems that it should be hard to
distinguish the contributions of interactions with the vector and
axial--vector baryon currents, caused by the right--handed electroweak
bosons like those in the electroweak models with left--right symmetries
\cite{Beg:1977ti,Holstein:1977qn,Carnoy:1988uv}, from the interactions
with the left(right)--handed scalar and tensor baryon currents.

The available experimental data on the correlation coefficients and
the lifetime of the neutron $\beta^-$--decay \cite{Ivanov:2012qe}
cannot be used for an estimate of the contributions of the scalar and
tensor interactions beyond the SM. As has been shown in
\cite{Ivanov:2012qe} the electron--neutron spin asymmetry
$A_{\exp}(E_e)$ and the proton recoil energy spectrum $a(T_p)$, where
$T_p$ is a kinetic energy of the decay proton, can be used for
measurements of the axial coupling constant $\lambda_{\rm eff}$ and
the Fierz term $b_F$. Then, the antineutrino--neutron spin asymmetry
$B_{\exp}(E_e)$ and the proton--recoil--neutron spin asymmetry
$C_{\exp}$ can give an information about the real parts of the scalar
$g_S$ and tensor $g_T$ coupling constants. In turn, the experimental
data on the lifetime of the neutron $(\tau_n)_{\exp}$, taken together
with the experimental data on the axial coupling constant
$\lambda_{\rm eff}$ and the Fierz term $b_F$ measured from the
electron--neutron spin asymmetry $A^W(E_e)$ and the proton recoil
energy spectrum $a(T_p)$, are able to give the information about the
CKM matrix element $V_{ud}$. Since available experimental data on the
electron--neutron spin asymmetry $A^W(E_e)$, the proton recoil energy
spectrum $a(T_p)$, the antineutrino--neutron spin asymmetry
$B_{\exp}(E_e)$ and the proton--recoil--neutron spin asymmetry
$C_{\exp}$ are not precise enough and were not elaborated together
with the aim to extract the experimental value of the axial coupling
constant and the experimental constraints on the scalar $g_S$ and
tensor coupling $g_T$ constants we propose to make an estimate of the
Fierz term $b_F$ by using the theoretical value of the neutron
lifetime $(\tau_n)_{\rm th} = 879.6(1.1)\,{\rm s}$, calculated in
\cite{Ivanov:2012qe}, and the world average value $(\tau_n)_{\exp} =
880.1(1.1)\,{\rm s}$ \cite{Beringer:1900zz}. Using the results,
obtained in \cite{Ivanov:2012qe} (see Chapter X of
Ref.\cite{Ivanov:2012qe}) we get
\begin{eqnarray}\label{label28}
b_F = - \frac{1}{\displaystyle \Big\langle
  \frac{m_e}{E_e}\Big\rangle}\,\Big(\frac{(\tau_n)_{\exp} -
  (\tau_n)_{\rm th}}{(\tau_n)_{\rm th}} \pm \frac{\sqrt{2}\, \Delta
  \tau_n}{~(\tau_n)_{\rm th}}\Big) = (- 0.87 \pm 2.70)\times 10^{-3},
\end{eqnarray}
where $\langle m_e/E_e\rangle = 0.6556$ \cite{Ivanov:2012qe} and
$\Delta \tau_n = 1.1\,{\rm s}$. At $\lambda_{\rm eff} = - 1.2750$ for
${\rm Re}(g_S + 3\lambda_{\rm eff} g_T)$ we get the following estimate
\begin{eqnarray}\label{label29}
{\rm Re}(g_S + 3\lambda_{\rm eff} g_T) = (- 2.56 \pm 7.93)\times
10^{-3}.
\end{eqnarray}
Since to order $10^{-3}$ the right-hand side (r.h.s) of
Eq.(\ref{label29}) is commensurable with zero, we may only assume that
${\rm Re}(g_S + 3\lambda_{\rm eff} g_T) \sim 10^{-4}$ or even
smaller. Hence, an expected value of the effective coupling constant
${\rm Re}(\lambda_{\rm eff} g_S - g_T)$ is ${\rm Re}(\lambda_{\rm eff}
g_S - g_T) \sim 10^{-4}$ . This results in that $10^{-5}$ is an
expected order of the contribution to the angular distribution $d
W^{(\pm)}_{F = 0}(\theta)/d\Omega$ of the term, proportional to ${\rm
  Re}(\lambda_{\rm eff} g_S - g_T) \sim 10^{-4}$, relative to the main
one. Thus, if the coupling constants of the right--handed leptonic
currents $|\bar{g}_S|$ and $|\bar{g}_T|$ are of order $|\bar{g}_S|\sim
|\bar{g}_T| \sim 10^{-4}$ the contribution of the left--handed
polarisation state of antineutrinos is of order $10^{-9}$ or even
smaller.

\section{Acknowledgement}

We are grateful to Hartmut Abele for fruitful discussions. This work
was supported by the Austrian ``Fonds zur F\"orderung der
Wissenschaftlichen Forschung'' (FWF) under the contracts I689-N16,
I534-N20 PERC and I862-N20.

\bibliographystyle{h-physrev3}
\bibliography{RevND2prc}

\end{document}